\documentclass{elsart}  

\usepackage{amssymb,amsmath,lineno,graphicx}

\begin{document}

\newcommand{\secref}[1]{$\S$\ref{#1}}
\newcommand{\figref}[1]{\figurename \ref{#1}}
\newcommand{\equref}[1]{Eq(\ref{#1})}
\newcommand{\tabref}[1]{Table \ref{#1}}

\newcommand{\Hline}{\rule{\linewidth}{0.25mm}}


\newcommand{\pwlw}{power-law }
\newcommand{\TPs}{TP statistic }
\newcommand{\pval}{{$p$-value}}

\newcommand{\temin}{{E_{\text{min}}}}
\newcommand{\teminH}{{\hat{E_{\text{min}}}}}
\newcommand{\temax}{{E_{\text{max}}}}
\newcommand{\teobs}{{E_{\text{obs}}}}
\newcommand{\tg}{{\gamma}}
\newcommand{\tgin}{{\gamma_{\text{IN}}}}
\newcommand{\tgout}{{\gamma_{\text{OUT}}}}
\newcommand{\tgH}{{\hat{\gamma}}}
\newcommand{\teb}{{E_{\text{b}}}}
\newcommand{\tebH}{{\hat{E_{\text{b}}}}}
\newcommand{\teh}{{E_{\frac{1}{2}}}}
\newcommand{\tehH}{{\hat{E_{\frac{1}{2}}}}}
\newcommand{\td}{{\delta}}
\newcommand{\tdH}{{\hat{\delta}}}
\newcommand{\tw}{{w_{c}}}
\newcommand{\twH}{{\hat{w_{c}}}}

\newcommand{\fM}{{f_{\text{M}}}}
\newcommand{\fP}{{f_{\text{P}}}}
\newcommand{\fDP}{{f_{\text{DP}}}}
\newcommand{\fFP}{{f_{\text{P}}}}
\newcommand{\fMt}{{\tilde{f}_{\text{M}}}}
\newcommand{\LM}{{\mathcal{L}_{\text{M}}}}
\newcommand{\LP}{{\mathcal{L}_{\text{P}}}}
\newcommand{\LDP}{{\mathcal{L}_{\text{DP}}}}
\newcommand{\LFP}{{\mathcal{L}_{\text{FP}}}}
\newcommand{\LMt}{{\tilde{\mathcal{L}}_{\text{M}}}}

\newcommand{\DKS}{{D_{\text{KS}}}}
\newcommand{\pKS}{{p_{\text{KS}}}}
\newcommand{\pTP}{{p_{\text{TP}}}}
\newcommand{\pR}{{p_{\mathcal{R}}}}


\begin{frontmatter}
\title{Statistical Tools for Analyzing the \\ 
       Cosmic Ray Energy Spectrum}

\author[New Mexico]{J.~D.~Hague\thanksref{email} }
\author[New Mexico]{B.~R.~Becker}
\author[New Mexico]{M.~S.~Gold}
\author[New Mexico]{J.A.J.~Matthews}

\address[New Mexico]{
  University of New Mexico, 
  Department of Physics and Astronomy, 
  Albuquerque, New Mexico, USA
}

\thanks[email]{Corresponding author, E-mail: \texttt{jhague@unm.edu} }
\date{\today}

\begin{abstract}  
In this paper un-binned statistical tools for analyzing the cosmic ray energy spectrum 
are developed and illustrated with a simulated data set. 
The methods are designed to extract accurate and precise model parameter 
estimators in the presence of statistical and systematic energy errors.
Two robust methods are used to test for the presence of flux suppression 
at the highest energies: the Tail-Power statistic and a likelihood ratio test. 
Both tests give evidence of flux suppression in the simulated data. 
The tools presented can be generalized for use on any astrophysical data set 
where the power-law assumption is relevant and can be used to aid observational design. 
\end{abstract}

\end{frontmatter}

\section{Introduction} \label{sec:Intro}
The observation of suppression in the flux of the highest energy cosmic rays 
(CRs) has been of central interest to astro-particle physics since the 
prediction of the GZK-effect\cite{refG,refZK} in 1966.
Most recently both the Auger\cite{Yamamoto:2007xj} and the HiRes\cite{Abbasi:2007sv}
detectors have released results favoring the observation of flux 
suppression at a $6\sigma$ and $5\sigma$ level of confidence, respectively.  

With this in mind, we describe a set of statistical tools 
designed to extract the most accurate and precise information 
concerning the flux of the highest energy cosmic rays. 
By binning the data we can only lose information\cite{refGolds} (see \secref{sec:BvUB})  
and therefore our statistical tools 
use an un-binned maximum likelihood approach\cite{refPDG, refHowell, refNewm, refClauset07} 
to answer two related statistical questions: 
{\it Is there flux suppression at the highest energies?} and, if yes, 
{\it What are the characteristic cut-off energy and shape parameters?}

In detail we first generate a toy data set 
using the CRPropa package\cite{refCRPropa}, as in \secref{sec:ats}. 
We then fit this simulated data to the three models described in \secref{sec:Models}.
The un-binned maximum likelihood fit is outlined in \secref{sec:ID} and 
methods for incorporating systematic and statistical energy errors are 
described in \secref{sec:SysEE} and \secref{sec:StatEE} respectively.
In \secref{sec:EtF} we describe several statistical tools for hypothesis testing: 
the Kolmogorov-Smirnov test, 
the tail power statistic\cite{refPisa1, refHague, Yamamoto:2007xj}, 
and a likelihood ratio test\cite{refHagueicrc1217}.

Though we cast our discussion in terms of cosmic ray energies, it is worth noting 
that these tools can be applied to any astrophysical data set where deviations 
from the power-law hypothesis are relevant, 
e.g. the galaxy correlation function\cite{refZehavi} or 
gamma ray astronomy\cite{refSchroedter}.

\section{CRPropa Data Set and Models} \label{sec:Data}

\subsection{Input from the HiRes and Auger Observatories} \label{sec:haa}
Both the HiRes\cite{Abbasi:2007sv} and Auger\cite{Yamamoto:2007xj} observatories have 
reported spectra and fit parameters for various power-law models. 
The collaborations use binned fitting methods.
They fit the spectrum over many orders of magnitude in energy but we summarize here 
the model parameters
\!\!\!\footnote{See \secref{sec:Models} and \tabref{tab:models} for the 
definition of these parameters.} 
relevant only to the highest energies.
The best fit double power-law parameters reported by HiRes\cite{Abbasi:2007sv} are 
$\tg = 2.81\pm0.03$(stat)$\pm0.02$(sys), 
$\teb = 10^{1.75\pm0.04}$(stat) and 
$\td = 5.1\pm0.7$(stat). 
For the same model Auger\cite{Yamamoto:2007xj} reports
$\tg = 2.62\pm0.03$(stat)$\pm0.02$(sys), 
$\teb = 10^{1.6}$(fixed) and 
$\td = 4.14\pm0.42$(stat). 
Fitting to the Fermi power-law Auger\cite{Yamamoto:2007xj} finds 
$\tg = 2.56\pm0.06$(stat), 
$\teh = 10^{1.74\pm0.06}$(stat) and 
$\tw = 0.16\pm0.04$(stat).
\begin{figure}[htbp] 
\Hline
\begin{center}
\includegraphics*[width=110mm, height=60mm]{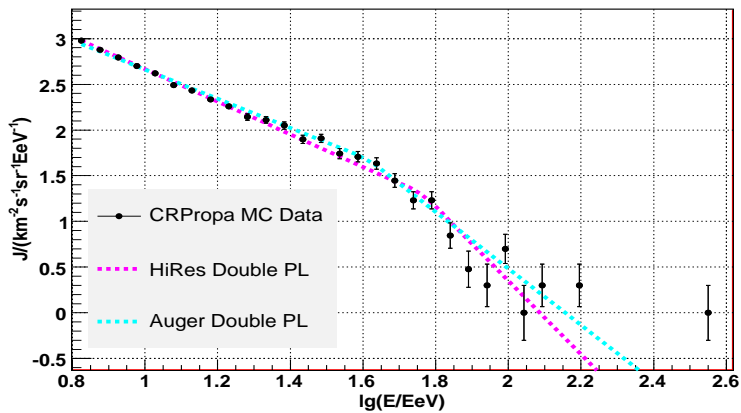} 
\end{center} 
\caption{\label{fig:pdfObs}
  The differential flux as simulated by $5\times10^{3}$ events from the CRPropa toy 
  set with parameters $\tgin = 2.6$ and $\temax = 2000$ EeV (see \secref{sec:ats}). 
  The p.d.f. of the best fit double power-laws 
  reported by HiRes\cite{Abbasi:2007sv} and Auger\cite{Yamamoto:2007xj} are the 
  dashed lines.
}
\Hline
\end{figure} 

\subsection{A Toy CR Data Set} \label{sec:ats}
To illustrate the methods in this note we use un-binned proton primary cosmic ray, CR, arrival energies 
(in EeV$\equiv10^{18}$eV) 
as simulated by the package CRPropa\cite{refCRPropa} with 
input spectral index $\tgin = 2.6$, $\temin=10$ EeV and $\temax = 2000$ EeV. 
We draw $5\times10^{3}$ events to act as a {\it toy} 
data set from a modern CR detector.

The CRPropa toy data set is similar size and shape to the flux reported 
by these observatories but the results of this study do not, otherwise, reflect 
any information about any physical data set. 
The probability distribution function (p.d.f.) of the best fit double power-laws 
reported by HiRes\cite{Abbasi:2007sv} and Auger\cite{Yamamoto:2007xj} are 
shown in \figref{fig:pdfObs} along with the CRPropa toy data.

The CRPropa propagation simulation is implemented by first generating proton CR primaries with 
initial energies according to a power-law ``at the source,'' propagating them through a 
simulated Universe and then observing the final energy. 
The spacial extent of the sources is simulated as a uniform distribution of 
discrete sources on a grid with $10$ Mpc steps extending to a distance of $4.07$ Gpc, 
(from redshift $z=0.0$ to $z=2.73$). 
Nuclei traveling over many megaparsecs from these sources will suffer significant energy loss 
in an expanding Universe filled with the cosmic microwave background, CMB, radiation.
As a result, the highest energy flux is much less than one would expect from a power-law 
alone. This suppression is known as the GZK-effect\cite{refG, refZK}. 

\begin{figure}[htbp] 
\Hline
\begin{center}
\includegraphics*[width=110mm, height=60mm]{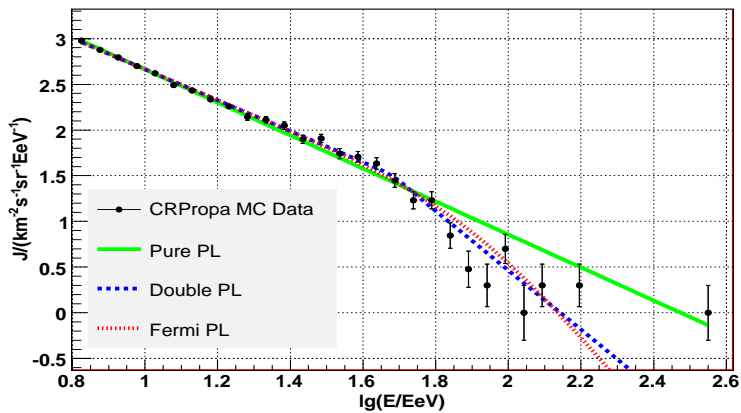} 
\end{center}
\caption{\label{fig:pdf}
  The differential flux as simulated by $5\times10^{3}$ events from the CRPropa toy 
  set with parameters $\tgin = 2.6$ and $\temax = 2000$ EeV. 
  The best fit models are described in \secref{sec:Models}.
}
\Hline
\end{figure}

\subsection{Power-Law Models} \label{sec:Models}
The fundamental probability distribution function governing 
the pure power-law assumption, denoted $\fP$, is shown in Table \ref{tab:models}: 
$f_{\text{P}} = (\tg-1)\temin^{\tg-1}E^{-\tg}$.
The parameter $\tg$ is referred to as the {\it spectral index}.
Here the sub-scripted-P stands for Pure-power-law.

For the highest energy CRs, the interesting observation would be to confirm 
or deny deviation from the power-law form at the highest magnitudes, i.e. the GZK-cutoff.
We therefore study two toy models that mimic a pure power-law for lower energies 
but exhibit flux suppression above a given energy.
The first is a double power-law (DP) with two spectral indexes, $\tg$ below
$\teb$ (``b'' for bend or break) and $\td > \tg$ above.
The point at which this p.d.f. reaches half the value it would have if
the pure power-law continued above $\teb$ is given by 
$\teh^{\text{dp}} = 2^{\frac{1}{\td-\tg}} \teb$, see \cite{Berezinsky88} for a discussion 
of this quantity. 
Both HiRes\cite{Abbasi:2007sv} and Auger\cite{Yamamoto:2007xj} have 
analyzed their data using this model.

We also study a toy p.d.f. where the cut-off is a ``Fermi-like'' Power-law 
(FP)\cite{Yamamoto:2007xj,refHague}.
The advantage of fitting with this toy model is that the location parameter 
$\teh$ is a parameter in the fit.

All three p.d.f.'s are normalized on the interval $[\temin, \infty)$, i.e. \\
$\langle \rangle_{\text{M}} \equiv \int_{\temin}^{\infty} f_{M}(t)dt=1$ 
for each of the models $\text{M} \in \{\text{P}, \text{DP}, \text{FP}\}$. 
The first element of the parameter vector $\theta_{1} \equiv \temin$ 
is fixed for the fit (see \secref{sec:Fit}) and then varied to estimate the 
stability (see \secref{sec:GoF}). 
Thus the power-law has one free parameter 
and the other models have three; low energy spectral index, location of cut-off and 
``steepness'' of cut-off.
\begin{table}[htbp]
\begin{center}
\begin{tabular}{|c|c|c|c|}
\hline
Model &$N_{\text{dof}}$& Normalization & Function  \\
\hline
P &1& $(\tg-1)\temin^{\tg-1}$& $E^{-\tg}$ \\
\hline
DP &3&  
$\frac{\tg-1}{\teb} 
\left\{\left(\frac{\teb}{\temin}\right)^{\tg-1} + 
             \frac{\tg-1}{\td-1} - 1 \right\}^{-1}$ & 
\begin{tabular}{c c}
$\left(\frac{E}{\teb}\right)^{-\tg}$ &$\temin \leq E < \teb$ \\
$\left(\frac{E}{\teb}\right)^{-\td}$ & $\teb \leq E$
\end{tabular} \\
\hline
FP &3& 
$\langle \rangle_{\text{FP}}^{-1}$, numerically&  
$E^{-\tg}\left[ 1 + \left(\frac{E}{\teh}\right)^{1/(\tw \ln 10)}\right]^{-1}$ \\
\hline
\end{tabular}
\end{center}
\caption{\label{tab:models}
  The model designation 
  (Model = Pure power-law, Double Power-law or Fermi Power-law), 
  number of free parameters, normalization, and form of the 
  function used to fit the simulated fluxes used in this study.
}
\Hline
\end{table}

\section{Fitting the Data} \label{sec:Fit}
We take an un-binned maximum log-likelihood approach to estimating 
the best-fit parameters of each model. 
The method constructed here is designed to 
extract the maximum possible statistical information about these parameters. 
For the ideal detector we assume that the observed energies are known 
with infinite precision. 

\subsection{Ideal Detector} \label{sec:ID}
We find estimates of the parameters in each model by maximizing,
\begin{linenomath*}
\begin{equation} \label{equ:lnQ}
\LM(\vec{\theta}) = \sum_{i=1}^{N} 
   \ln \left\{ \fM(E_{i}; \vec{\theta}) \right\}, 
\end{equation}
\end{linenomath*}
where the sum is carried out over the event energies and 
$\theta_{1} \equiv \temin$ is fixed.
The global maximum of this
function $\LM(\hat{\theta})$ determines the best parameter 
estimates, $\hat{\theta}$. 
The the function is maximized using Minuit\cite{refMINUIT}
with the MIGrad option. 

To determine the one degree of freedom error estimate\cite{refPDG} for a parameter 
we vary the parameter (with the others fixed at $\hat{\theta}$) 
until $-2\Delta \LM = 1$. 
The two degrees of freedom error estimates\cite{refPDG} are determined by varying 
two parameters with the other fixed and choosing the contour such that 
$-2\Delta\LM \geq 2.3$.
For the toy data set, we plot these contours and the asymmetric one 
degree of freedom error estimates 
in \secref{sec:SumToy}: \figref{fig:dplLL2} and \ref{fig:fplLL2}.

\subsection{Systematic Energy Error} \label{sec:SysEE}
The errors on the observed energy $E_{obs}$ of an event from a real
CR detector are considerable and must be included in any realistic analysis of a spectrum. 
For our purposes, these errors take the two canonical forms; 
{\it statistical} and {\it systematic}, 
i.e. $E_{obs}\pm\sigma_{stat}\pm\sigma_{sys}$.

The systematic errors energy errors of a CR detector 
reflect the uncertainties in the absolute calibration of the detector. 
At the highest energies the 
systematics are the dominant contribution to the overall uncertainty 
of an event's energy. 
For example, the two fluorescence detectors Auger\cite{Yamamoto:2007xj} and 
Hires\cite{Abbasi:2007sv} report uncertainties of 22\% and 17\% respectively.
\!\!\!\footnote{With its hybrid detector the Auger reduces the systematic error to 
between 7\% and 15\%\cite{Yamamoto:2007xj}.}
The shift in energy due to the systematic error can be asymmetric, 
i.e. $\sigma^{+}_{sys} \neq \sigma^{-}_{sys}$, 
and energy dependent, see \equref{equ:sigE},
but it effects every event at a given energy the same way; a shift up or down. 
For the Monte-Carlo (MC) data sets we model the systematic detector energy errors using:
\begin{linenomath*}
\begin{equation} \label{equ:sigE}
\frac{\sigma(E ; \vec{p})}{E} = p_{1} + p_{2}\lg(E).
\end{equation}
\end{linenomath*}
Here we choose symmetric systematically-shifted energies such that 
the energy of the $k^{th}$ event 
is $E^{\pm}_{k} = E_{k} \pm  \sigma(E_{k} ; \vec{p}_{sys})$. 
For the systematic errors we choose $p_{1}=0.05$ and $p_{2}=0.10$. 

To account for this in the parameter estimation procedure, we 
shift each energy up or down and carry out the methods in \secref{sec:ID}. 
The difference between the parameter estimates of a shifted set 
and those of the centered set gives ``systematic'' errors of the parameter 
estimates.

\subsection{Statistical Energy Error} \label{sec:StatEE}
To model the statistical energy errors of the detector we assume that 
the {\it true} energy of the cosmic ray has a $68\%$ chance of being 
within the interval $(E_{obs}-\sigma_{stat}, E_{obs}+\sigma_{stat})$. 
The observed energy has been ``smeared'' from the true value; 
$E_{obs} = E_{true} + Y$ where $Y$ is drawn from 
a normal distribution with mean $0$ and variance $\sigma_{stat}$.
Note that while the true energies can only be found on $[\temin,\infty)$, there 
is a nonzero probability for the (after smearing) observed energy to be less than 
$\temin$; $E_{obs}$ lives on the interval $(-\infty, \infty)$.
This edge effect near $\temin$ can be 
accounted for by assuming that the true distribution of energies 
follows a power-law well below $\temin$ and then re-normalizing the 
convolution technique used in Howell\cite{refHowell}.
See \secref{sec:MCexe} for further discussion. 
For the integrand, three factors are necessary: 
\begin{enumerate}
\item The model to be fitted, $\fM(t; \vec{\theta})$ (see \secref{sec:Models}).
  By letting $\theta_{0}=0.1\temin$ we are assuming that the power-law extends 
  below the observed $\temin$.
\item A normal distribution $G(t; \teobs, \sigma_{stat}(t;\vec{p}))$ with 
  mean $\teobs$ and variance $\sigma_{stat}(t;\vec{p})$ to reflect the statistical 
  energy errors.
\item The acceptance of the CR detector as a function of the true energies $\Omega(t)$. 
  Since we are using MC data we choose $\Omega(t) = 1$ for simplicity.
\end{enumerate}
The convolution is calculated by integrating over all possible 
{\it true} energies ($t$):
\begin{linenomath*}
\begin{equation} \label{equ:gM}
g_{\text{M}}(\teobs; \vec{\theta}, \vec{p}) = \int^{\infty}_{0.1\temin} 
                        \fM(t; \vec{\theta}) \,
                        G(t; \teobs, \sigma_{stat}(t;\vec{p}))\, \Omega(t) \, dt.
\end{equation}
\end{linenomath*}
Re-normalizing so that the observed energies define a p.d.f., 
we numerically calculate the p.d.f. to be: 
\begin{linenomath*}
\begin{equation} \label{equ:fMt}
\fMt(\teobs; \vec{\theta}, \vec{p}) = \frac{g_{\text{M}}(\teobs; \vec{\theta}, \vec{p})}
    {\int^{\infty}_{\temin} g_{\text{M}}(y; \vec{\theta}, \vec{p})\, dy}, 
\end{equation}
\end{linenomath*}
and we must modify the likelihood found in \equref{equ:lnQ} accordingly:
\begin{linenomath*}
\begin{equation} \label{equ:lnQt}
\LMt(\vec{\theta}) = \sum_{i=1}^{N} \ln \left\{ \fMt(E_{i}; \vec{\theta}) \right\}.
\end{equation}
\end{linenomath*}
By finding the parameters $\hat{\theta}$ which 
maximize \equref{equ:lnQt} we can be confident that we are accounting for the 
statistical uncertainty inherent in data collected by a realistic detector.
To model statistical errors in our toy data set, we parameterize $\sigma_{stat}$ 
as in \equref{equ:sigE} with $p_{1}=0.15$ and $p_{2}=0$.

\section{Evaluating the Fit} \label{sec:EtF}
In this section we outline ways to evaluate the fit of a candidate model to the data set. 
The Kolmogorov-Smirnov statistic can be used to extract a best fit minimum energy $\hat{\temin}$ 
and, with its corresponding \pval, evaluate the ``absolute goodness of fit'' 
of a candidate model (see \secref{sec:GoF}). 
The relevant question for CR physics is not whether a particular model is a good fit to 
the data but rather whether the flux exhibits suppression (relative to the 
single power-law form) at the highest energies. 
To address this question directly we use two statistics with well 
defined $p$-values: the Tail-Power statistic (see \secref{sec:TP}), which can give information 
about tail suppression in standard deviations, 
and a likelihood ratio that allows rejection of 
the single power-law hypothesis in favor of a suppressed candidate model 
(see \secref{sec:MD}). 

\subsection{Kolmogorov Statistic} \label{sec:GoF}
While the minimum value of the likelihood function will indeed give the best value 
of the fit parameters, this fit may nonetheless be poor. 
The typical\cite{refGolds,refClauset07} method for evaluating goodness of fit is the 
Kolmogorov-Smirnov test\cite{refPDG}.
The relevant statistic for this test is the KS distance:
\begin{linenomath*}
\begin{equation} \label{equ:Ks}
\DKS(\temin) = \max_{E\geq \temin} \left| F_{\text{fit}}(E) - 
                                    F_{\text{data}}(E) \right|, 
\end{equation}
\end{linenomath*}
where, $F_{\text{fit}}$ and $ F_{\text{data}}$ are the cumulative distribution functions 
(c.d.f.) of the best fit model and the data respectively. The maximum 
distance between the c.d.f.'s is taken over all energies in the fitted data set, 
$E\geq \temin$.
By stepping over $\temin$ and re-minimizing \equref{equ:lnQ} at each step to determine 
the best fit parameters, we can calculate $\DKS$ as a function of $\temin$.
The value of $\hat{\theta}_{0} \equiv \teminH$ that minimizes $\DKS$ can be taken 
as the best estimate of the minimum energy above which the model holds\cite{refClauset07}.

To test how well a particular model fits the data we must simulate many MC data 
sets drawn from the best fit model p.d.f. with the same number of events as the 
original data. 
The fraction of sets $\pKS$ with $\DKS$ greater than that of the data gives 
the suitable $p$-value; if $\pKS \ll 1$ then it is unlikely that the data are 
drawn from the model under consideration, 
and in this way the KS test statistic $\pKS$ can rule out the different candidate 
models\cite{refClauset07}. 

\subsection{Tail Power Statistic} \label{sec:TP}
The Tail-Power (TP) statistic is similar to the KS statistic discussed above, 
however it has, at least, three advantages over $\pKS$ when testing the 
power-law assumption;
\begin{enumerate}
\item The \TPs and it's corresponding $p$-value $\pTP$ are 
  nearly independent of the value of the spectral index $\tg$,
\item The asymptotic behavior of the \TPs is known, and therefore no simulations 
  are required to calculate the corresponding $p$-value $\pTP$, 
\item If $\text{TP}>0$ the deviation suggests flux {\it suppression in the tail} and 
  if $\text{TP}<0$ the deviation suggests flux 
  {\it  enhancement in the tail}\cite{refHague} and 
\item $\pTP$ offers an unambiguous $p$-value in standard deviations.
\end{enumerate}
This ``measure of power-law-ness'' has been developed and studied elsewhere
(see \cite{refPisa1,Yamamoto:2007xj,refHague}) 
and here we expand its use to the un-binned case. 

The sample \TPs is defined as \cite{refPisa1}: 
\begin{linenomath*}
\begin{equation} \label{equ:TP}
\hat{\tau}(\temin) = \hat{\nu}^{2}_{1}(\temin) - \frac{1}{2}\hat{\nu}_{2}(\temin), 
\end{equation}
\end{linenomath*}
where:
\begin{linenomath*}
\begin{equation} \label{equ:nu}
\hat{\nu}_{n}(\temin) = \frac{1}{N_{>}} \sum_{E_{i}>\temin} \ln^{n} \frac{E_{i}}{\temin}
\end{equation}
\end{linenomath*}
and the sum is carried out over all $N_{>}$ events with energy greater than a given minimum.
If the data are drawn from a pure power-law then $\hat{\tau}(\temin)$ will tend 
to zero as $N \rightarrow \infty$, regardless of the value of $\tg$\cite{refGolds}. 

We may approximate the asymptotic joint distribution of ${\hat \nu}_{1}$ and ${\hat \nu}_{2}$ as 
a bivariate Gaussian $f_{{\scriptscriptstyle \nu_{1}\nu_{2}}}(\nu_{1}, \nu_{2})$. 
The asymptotic mean and variance of $\nu_{1}$ are $\frac{1}{\tg-1}$ 
and $\frac{1}{N (\tg-1)^{2}}$, and of $\nu_{2}$ are 
$\frac{2}{(\tg-1)^{2}}$ and $\frac{20}{N (\tg-1)^{4}}$. The random variables  
$\nu_{1}$ and $\nu_{2}$ are highly correlated; the correlation 
coefficient is  $\rho = \frac{2}{\sqrt{5}}$, {\it independent of $\tg$}.
Thus, for a given $N$ and $\tg$, we calculate the p.d.f. of $\tau$ to be,
\begin{linenomath*}
\begin{equation}\label{equ:pdftau}
f_{{\scriptscriptstyle TP}}(\tau; N, \tg) = \int^{\infty}_{-\infty} 
                                            f_{\nu_{1}\nu_{2}}(t, 2(t^{2}-\tau))dt.
\end{equation}
\end{linenomath*}
The analytic ``location'' $\langle \tau \rangle_{{\scriptscriptstyle TP}} \sim 0$ 
and ``shape'' 
$\langle \sigma_{\tau} \rangle_{{\scriptscriptstyle TP}} = 
\sqrt{ \langle\tau^{2}\rangle_{{\scriptscriptstyle TP}}- 
\langle \tau \rangle^{2}_{{\scriptscriptstyle TP}} } 
\sim N^{-1/2}(\tg-1)^{-2}$ 
parameters of this distribution are consistent with simulation generated values.
We measure the $p$-value $\pTP$ for the TP statistic in units of 
standardized deviation,
\begin{linenomath*}
\begin{equation}\label{equ:pTP}
\pTP(\temin) = \frac{\hat{\tau}(\temin) - \langle \tau \rangle_{{\scriptscriptstyle TP}}}
                       {\langle \sigma_{\tau} \rangle_{{\scriptscriptstyle TP}}}.
\end{equation}
\end{linenomath*}
A spectrum with flux suppression in the tail (like that in the Fermi-like model) 
will result in a {\it positive} significance\cite{refHague}. 

\begin{figure}[htbp] 
\Hline
\begin{center}
\includegraphics*[width=120mm, height=70mm]{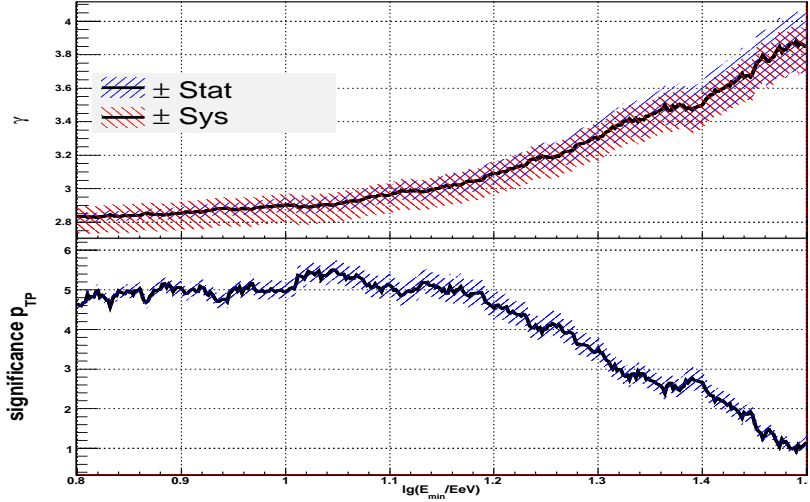} 
\end{center}
\caption{\label{fig:pplestToy}
  {\it Top} The best fit (see \secref{sec:ID}, \equref{equ:lnQ}) 
  spectral index $\tgH$ as a function 
  of $\lg \temin$ for the the toy CR data set(see \secref{sec:ats}) 
  fit to the pure power-law model (P). 
  {\it Bottom} The resulting \TPs significance $\pTP(\temin)$ in standard 
  deviations as a function of the minimum energy $\temin$, see \equref{equ:pTP}.
  Both plots give strong evidence of flux suppression of the highest energy 
  MC events.
}
\Hline
\end{figure}

The application of \equref{equ:pTP} to the toy CR data set (see \secref{sec:ats}) 
is plotted in \figref{fig:pplestToy}. 
The top panel shows the (pure power-law) spectral index as a function of 
$\temin$. 
A spectral index which increases as $\temin$ increases is indicative of flux suppression. 
The red, left leaning hatching shows the variation of $\tgH$ due to a 
$\pm1\sigma$ {\it systematic} shift in the energies (see \secref{sec:SysEE}) while the 
opposite, blue hatching shows the {\it statistical} error of the estimator $\tgH$, see 
\secref{sec:ID}.
The bottom panel shows the resulting \TPs significance $\pTP(\temin)$ in standard 
deviations. 
Notice that while the systematic errors can be significant for the 
measured spectral index, they do not effect the TP statistic. 
Since we must estimate the spectral index to compute $\pTP$, we also 
propagate the statistical errors on $\tgH$ to the tail power statistic.

\begin{figure}[htbp] 
\Hline
\begin{center}
\begin{tabular}{l l}
\includegraphics*[width=75mm, height=55mm]{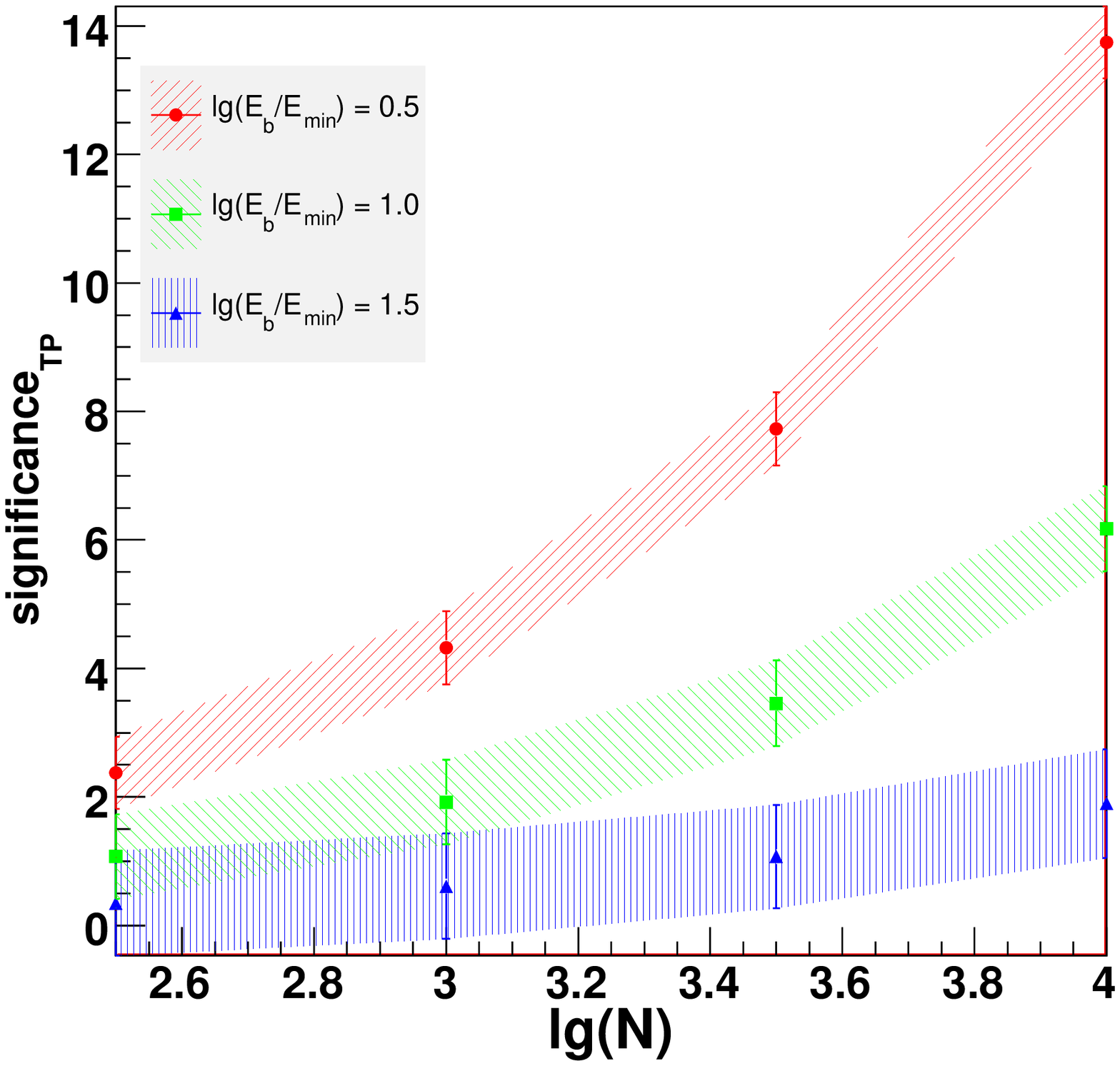} 
\includegraphics*[width=75mm, height=55mm]{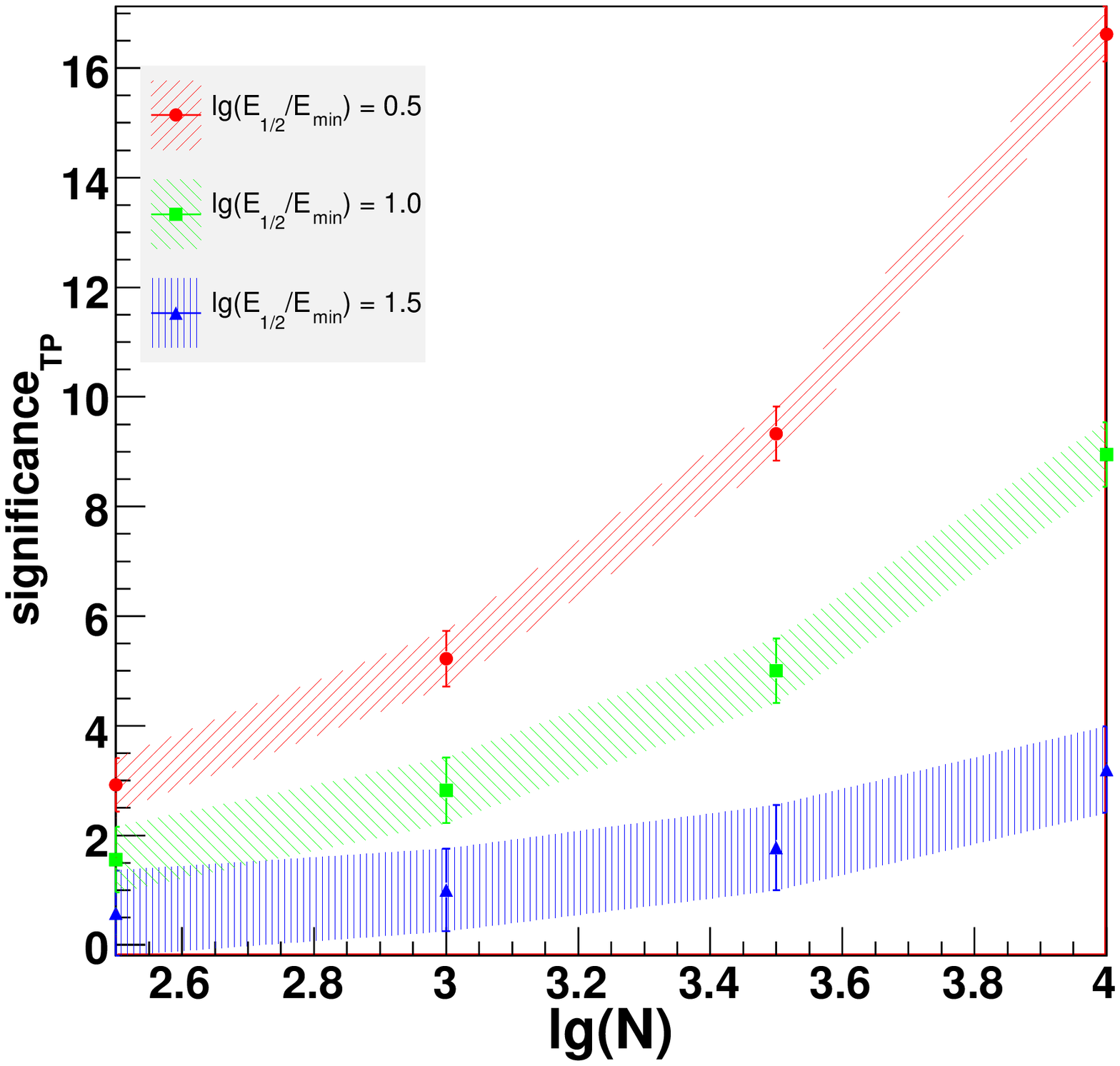} 
\end{tabular}
\end{center}
\caption{\label{fig:TPstats}
  The tail power significance, $\pTP(\temin=1.0)$ as a function of the 
  ($\log_{10}$ of the) number of events in each Monte-Carlo realization. 
  Each plot style represents a different choice of 
  $\lg(E_{\text{cut}}/\temin) = 0.5,\,1.0,\,\text{or}\,1.5$. 
  {\it Left}, the double power-law, $E_{\text{cut}} \equiv \teb$.
  {\it Right}, the Fermi power-law, $E_{\text{cut}} \equiv \teh$.
}
\Hline
\end{figure}

To test the effectiveness of this statistic, we apply it to a series of simulated data 
sets drawn from both the Fermi and double power-law models. 
For all the models we set
\!\!\!\footnote{These values are similar to the Auger\cite{Yamamoto:2007xj} and 
HiRes\cite{Abbasi:2007sv} best fit values.} 
$\temin = 1.0$EeV, $\tg=2.75$ and either $\td=4.75$ or $\tw=0.10$. 
We vary each characteristic cut-off energy, either $\teb$ or $\teh$, 
in three steps $\lg(E_{\text{cut}}/\temin) = 0.5,\,1.0,\,\text{and}\,1.5$. 
The total number of events in the data set is varied in four steps 
$\lg(N) \sim 2.5,\,3.0,\,3.5,\,4.0$. 
For each of these twelve sets of parameter choices we make $10^{3}$ Monte-Carlo 
realizations and plot the mean and RMS of $\pTP(\temin=1.0)$ in \figref{fig:TPstats}. 

Based on \figref{fig:TPstats} we can see that the best way to evaluate a data set 
with a potential for tail suppression is to collect as much data with $\temin$ as close 
to the expected cut-off as possible.
The experimenter may use \figref{fig:TPstats}, or one like it, to help tune 
observation parameters, i.e. collecting time on a gamma ray source or 
size of a CR detector, in advance of the observation and in anticipation of flux suppression 
of a certain type.
Note, however, that one should choose an $\temin$ {\it prior} to analyzing a data set to 
avoid a penalty for scanning in this parameter.

\subsection{Model Discrimination} \label{sec:MD}
Here we introduce a likelihood ratio test designed to discriminate 
candidate suppressed models (DP and FP) from the pure power-law.
We define two log-likelihood ratios; for each model M:
\begin{linenomath*}
\begin{equation} \label{equ:R}
\mathcal{R}_{\text{M}} = \sum^{N}_{i=1} \left\{ \ell_{\text{M}}(E_{i}) - 
                                      \ell_{\text{P}}(E_{i}) \right\}
                      = \LM - \LP,
\end{equation}
\end{linenomath*}
where $\ell_{\text{M}}(E_{i}) = \ln \fM(E_{i}; \hat{\theta})$ 
with M either DP (double power-law) or FP (Fermi-like), and
$\ell_{\text{P}}(E_{i}) = \ln \fP(E_{i}; \hat{\theta})$ for the pure 
power-law likelihood per event (see Table \ref{tab:models} and \equref{equ:lnQ}).
Note that each suppressed model is fit {\it independently} of the pure power-law 
best fit. The asymptotic variance of $\mathcal{R}$ can be estimated by the sample value:
\begin{linenomath*}
\begin{equation} \label{equ:sigR}
\sigma^{2}_{\mathcal{R}} = \frac{1}{N} \sum^{N}_{i=1}\left\{ 
                             [\ell_{\text{M}}(E_{i}) - \ell_{\text{P}}(E_{i})] - 
                             \left[\frac{\LM-\LP}{N}\right]
                          \right\}^{2},
\end{equation}
\end{linenomath*}

The hypothesis of the pure power-law is {\it nested} within the 
hypothesis of a suppressed power-law. As a consequence, 
$|\mathcal{R}|/\sigma_{\mathcal{R}} \rightarrow 0/0$ as $N \rightarrow \infty$ 
and the distribution of $\mathcal{R}/\sigma_{\mathcal{R}}$ is not Gaussian\cite{refClauset07}. 
The correct $p$-value is calculated
as the integral of a $\chi^{2}$ function\cite{refVuong, refClauset07}: 
\begin{linenomath*}
\begin{equation} \label{equ:pR}
p_{\mathcal{R}}(z^{2}) = \frac{1}{\sqrt{2\pi}}\int^{\infty}_{z^{2}} t^{-1/2}e^{-t/2}dt,
\end{equation}
\end{linenomath*}
where $z^{2} = \mathcal{R}^{2}_{\text{M}} / \left(2N\sigma^{2}_{\mathcal{R}}\right)$. 

\begin{figure}[htbp] 
\Hline 
\begin{center}
\begin{tabular}{c c}
\includegraphics*[width=75mm, height=55mm]{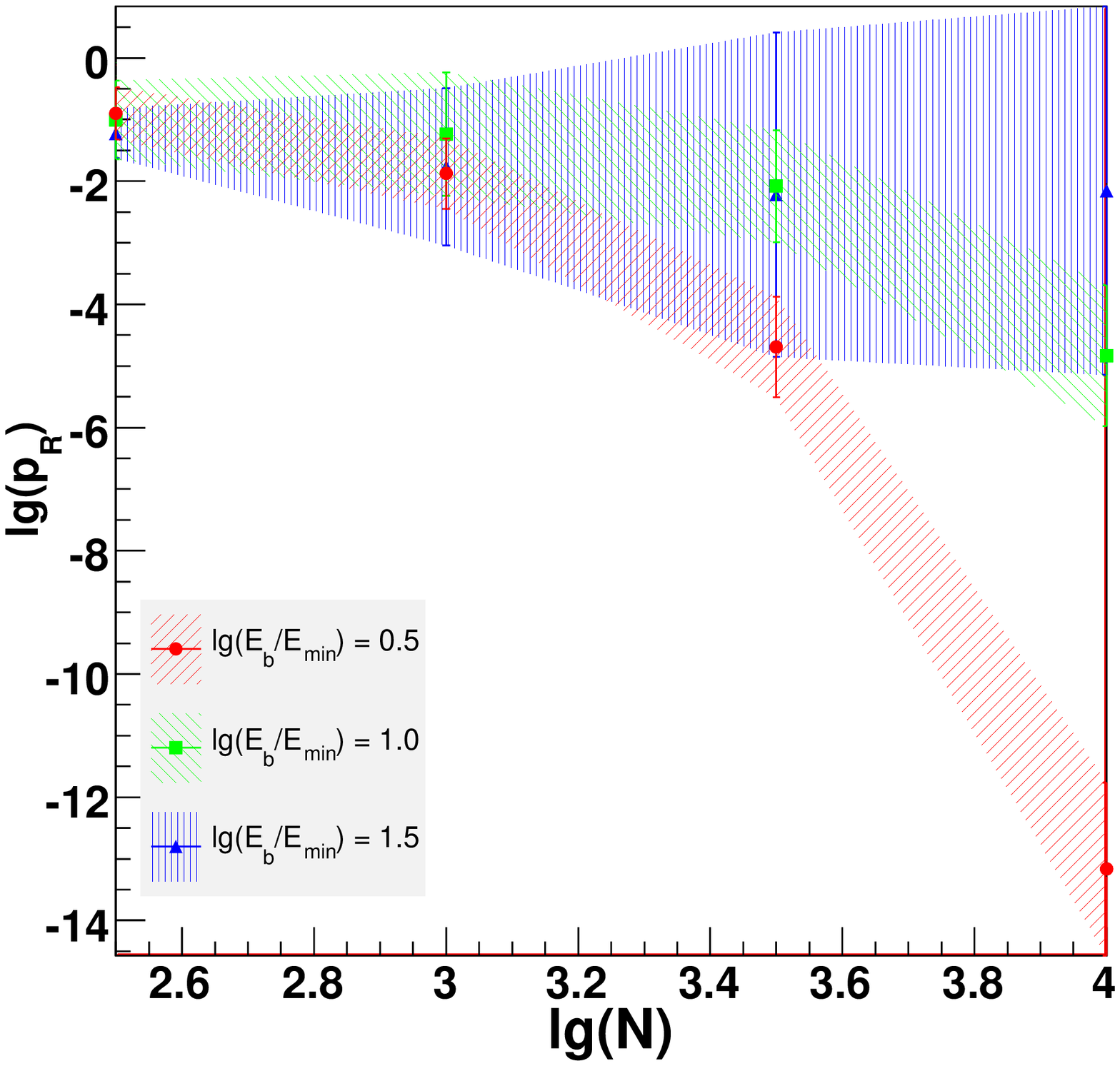} 
\includegraphics*[width=75mm, height=55mm]{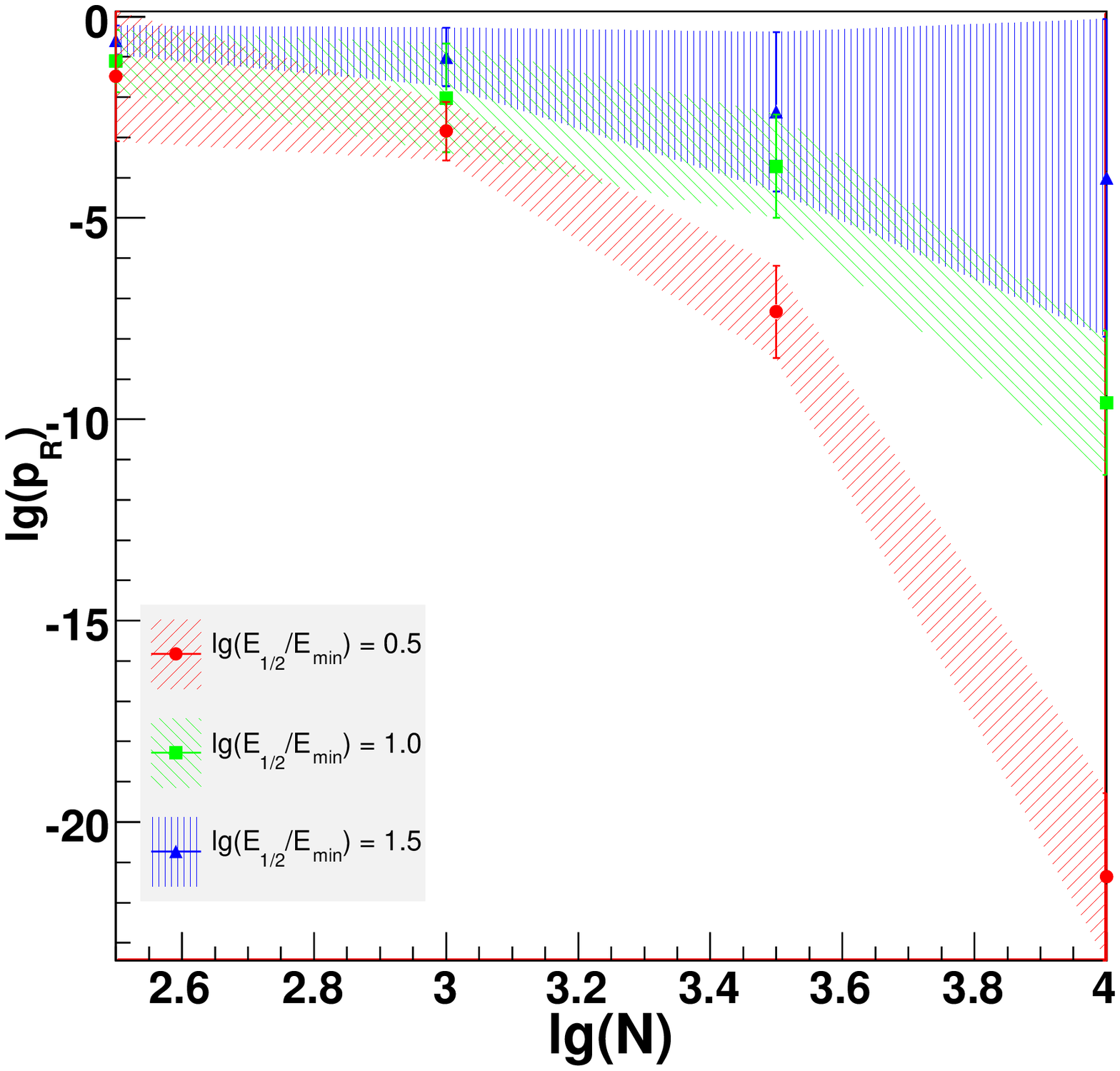} 
\end{tabular}
\end{center}
\caption{\label{fig:LRstats}
  The log of the likelihood ratio significance, $\pR$ as a function of the 
  ($\log_{10}$ of the) number of events in each Monte-Carlo realization. 
  Each plot style represents a different choice of 
  $\lg(E_{\text{cut}}/\temin) = 0.5,\,1.0,\,\text{or}\,1.5$. 
  {\it Left}, the double power-law, $E_{\text{cut}} \equiv \teb$.
  {\it Right}, the Fermi power-law, $E_{\text{cut}} \equiv \teh$.
}
\Hline
\end{figure}

We interpret this $p$-value in the following way: if $p_{\mathcal{R}}$ is ``small'' 
then the best fit model M may be preferred over the best fit pure power-law. 
By small we mean that, {\it a priori} and rather arbitrarily, we may choose to reject 
the single power-law in favor of the 
model if $p_{\mathcal{R}} \leq 10^{-3}$. This quantity tells us only whether a 
given suppressed model is better than the pure power-law. It says nothing 
about how well any of the fits actually represent the data.

For each of the twelve sets of parameter choices used in \figref{fig:TPstats}, 
we plot the mean and RMS of $\pR$ in \figref{fig:LRstats}. 
As before, we see that the best way to reject the power-law in favor of the suppressed model 
is to collect as much data with $\temin$ as close to the expected cut-off as possible. 
Note that for $\lg(E_{\text{cut}}/\temin) = 1.5$ the distribution of likelihood ratios 
is strictly positive and highly peaked near zero; the mean and RMS are not good reflections 
of this distribution.

\section{Summary and Conclusion}\label{sec:SaC}
In this paper we describe a set of statistical tools designed to extract the 
most accurate and precise information about the flux of the highest energy cosmic rays. 
We show how to use the un-binned likelihood method described in \secref{sec:ID} to 
fit a data set to the three model distributions described in \secref{sec:Models}. 
Techniques for incorporating the systematic and statistical errors associated with 
a real CR detector into the likelihood method are described in 
\secref{sec:SysEE} and \secref{sec:StatEE} respectively. 
In \secref{sec:EtF} we describe $p$-values useful for extracting information 
about flux suppression. 
We show in \secref{sec:TP} and \secref{sec:MD} how an experimenter might use an 
{\it a priori} estimate of the cut-off energy to maximize an observational setup 
for detecting flux suppression.

The collection of these statistical tools are the primary result of this paper. 
To answer the questions posed in the introduction for a given data set we 
suggest the following steps:
\begin{enumerate}
  \item Estimate the best fit parameters $\hat{\theta}$ of the model; \label{step:BF}
    \begin{enumerate}
    \item The estimates $\tgH$, $\tebH$ or $\tehH$ and $\tdH$ or $\twH$ are 
      determined via the likelihood \equref{equ:lnQ},\label{step:BFa}
    \item The estimate of the minimum energy $\teminH$ is that which minimizes the 
      Kolmogorov distance $\DKS$ (see \secref{sec:GoF}).\label{step:BFb}
    \end{enumerate}
  \item Shift the energies up and down according to the systematic uncertainty 
    described in \secref{sec:SysEE} and repeat step (\ref{step:BF}). 
    The resulting shift in parameter estimates gives the systematic uncertainty 
    of those estimates. \label{step:SysE}
  \item Obtain the model parameter estimates using the methods in \secref{sec:StatEE} 
    to incorporate the statistical error of each event energy. \label{step:StatE}
  \item Test the model hypothesis; \label{step:GoF}
    \begin{enumerate}
    \item The absolute goodness of fit for any of the models can be evaluated using 
      $\pKS$ in \secref{sec:GoF},\label{step:GoFa}
    \item The Tail-Power statistic $\pTP$ can be used to reject the single power-law 
      hypothesis (nearly independently of the spectral index estimate, 
      see \secref{sec:TP})\label{step:GoFb}
    \item The single power-law may be rejected in favor of a specific alternative 
      model using $\pR$, here we study the double and Fermi power-law distributions 
      (see \secref{sec:MD}).\label{step:GoFc}
    \end{enumerate}    
\end{enumerate}
The best estimates for the {\it characteristic cut-off energy and shape parameters}, 
determined via steps (\ref{step:BF}), (\ref{step:SysE}) and (\ref{step:StatE}), 
are $\tebH$ or $\tehH$ and $\tdH$ or $\twH$ respectively. 
The presence of {\it flux suppression at the highest energies} 
can be evaluated using step (\ref{step:GoF}). 

By applying these methods to the toy Monte-Carlo set of CRPropa events 
we illustrate in \secref{sec:SumToy} how the procedure may be implemented on an 
actual CR detector, i.e. a detector with systematic and statistical event energies. 
Suppression in the tail 
is clear in \figref{fig:cdf} and \figref{fig:rescdf}; the tail power statistic 
is $4.6\sigma$ and the $p$-value for the double (Fermi) power-law is 
$\lg p_{\text{DP}} = -2.7$ ($\lg p_{\text{FP}} = -1.9$).

The methods are sufficient and robust. 
Indeed, many of them have been applied by the Auger collaboration which 
reports suppression with $6\sigma$ confidence\cite{Yamamoto:2007xj}.
These tools serve as a basis for further investigation of the CR 
spectrum such as evidence for more detailed spectral information.
They can be applied to any data set, astrophysical or otherwise, to 
provide information both about data already collected and help to 
optimize future observations for detecting tail suppression. 

\newpage
\appendix
\section{Binned vs. Un-Binned} \label{sec:BvUB}
The statistical superiority of an un-binned maximum likelihood estimate of the 
pure power-law spectral index to the logarithmically binned least-$\chi^{2}$ method 
often used has been established 
in \cite{refGolds} and expanded upon more recently in 
\cite{refHowell,refNewm,refClauset07,refHague,refHagueicrc1217}.
In this section we compare the binned to the un-binned fitting method 
for the two suppressed models, i.e. the double and Fermi power-laws (see \secref{sec:Models}).

To calculate the binned estimators we minimize a $\chi^{2}(\vec{\theta})$ function 
that relates the logarithmically binned (width $w$) histogram of the data to that 
expected by a model.
The function is
\!\!\!\footnote{For the case of the single power-law $\lg \fP = \lg C - \tg \lg E$ where $C$ is the normalization. 
Thus the binned fitting method reduces to fitting the log$_{10}$ of the (error weighted) bin heights to a straight line with slope $\tg$. 
This technique is often used to mitigate the effects of the heaviness of the power-law tail but un-binned methods are more accurate and precise.},
\begin{linenomath*}
\begin{equation} \label{equ:chi2}
\chi^{2}(\vec{\theta}) = \sum_{i=1}^{N_{b}} \left( 
                       \frac{\lg Y_{i}^{\text{data}} - \lg Y_{i}^{\text{fit}}(\vec{\theta})}
                       {\sigma_{i}^{\text{data}}} \right)^{2}, 
\end{equation}
\end{linenomath*}
where $N_{b}$ is the number of bins, $Y_{i}^{\text{data}}$ is the 
number of events in the $i^{\text{th}}$ bin $b_i$ and 
$\sigma_i$ is determined by Gaussian errors when $Y_{i}^{\text{data}}>10$ and 
Poissonian errors when $Y_{i}^{\text{data}}\leq10$.
We minimize with respect to the parameters $\vec{\theta}$ 
(with $\theta_{0}\equiv\temin$ fixed) using the number of events in a bin 
expected by the model M, 
\begin{linenomath*}
\begin{equation} 
Y_{i}^{\text{fit}}(\vec{\theta}) =  N
                                 \int_{10^{b_{i}-w/2}}^{10^{b_{i}+w/2}} \fM(t; \vec{\theta}) dt
                                 . \nonumber
\end{equation}
\end{linenomath*}

To study the asymptotic bias and error produced by the two estimation techniques 
we draw $10^{5}$ sets of $5\times10^{3}$ events from a pure power-law 
and separately from a double distribution. 
For each Monte-Carlo set we estimate the best fit model parameters $\hat{\theta}$ 
using both the likelihood \equref{equ:lnQ} and the $\chi^{2}$ \equref{equ:chi2} methods.
The un-binned estimator of the pure power-law spectral index (see \secref{sec:ID}) 
has been shown\cite{refGolds, refHowell} to have an error estimate within $\sim 1\%$ 
of the Cramer-Rao lower bound for a sample with as few as $\sim 100$ events.

\begin{figure}[htbp] 
\Hline
\begin{center}
\includegraphics*[width=90mm, height=50mm]{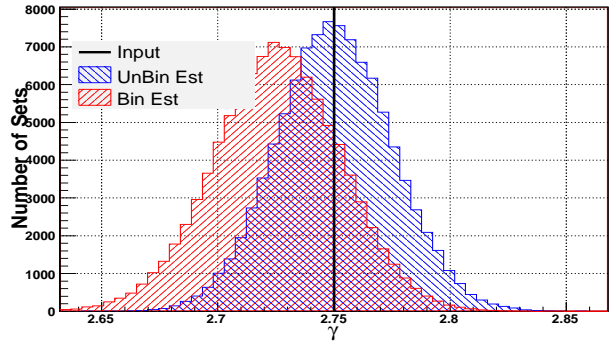} 
\end{center}
\caption{ \label{fig:pplbvub}
  For each of $10^{5}$ sets of $5\times10^3$ events drawn from a pure power-law with 
  index $\temin=1.0$ and $\tg=2.75$ 
  we estimate the spectral index using the binned \equref{equ:chi2} and un-binned 
  \equref{equ:lnQ} methods. 
  The bias and error of the un-binned estimator is $0.0002$ and $0.0247$ and that of the 
  binned is $-0.024$ and $0.0272$.
}
\Hline
\end{figure}

\begin{figure}[htbp] 
\Hline
\begin{center}
\includegraphics*[width=150mm, height=60mm]{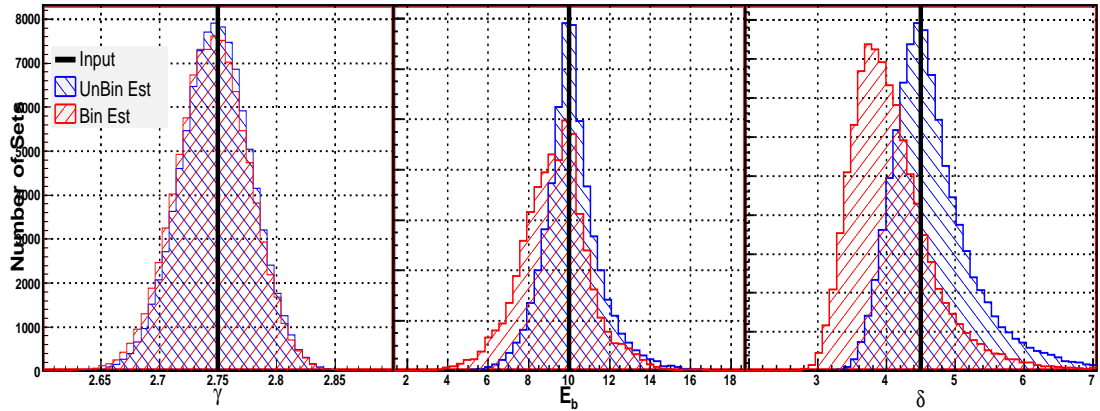} 
\end{center}
\caption{ \label{fig:dplbvub}
  For each of $10^5$ sets of $5\times10^3$ events drawn from a double power-law with 
  parameters $\{\tg, \teb, \td\} = \{2.75, 10.0, 4.5\}$ 
  we estimate the spectral index using the binned \equref{equ:chi2} and un-binned 
  \equref{equ:lnQ} methods. 
  The bias and error of the un-binned estimators are 
  $\{-0.002, 0.13, 0.16\}$ and $\{0.03, 1.4, 0.60\}$ and those of the 
  binned are $\{-0.005, -0.71, -0.43\}$ and $\{0.03, 1.7, 0.60\}$.
}
\Hline
\end{figure}

In \figref{fig:pplbvub} and \figref{fig:dplbvub} we plot the results of the 
simulations. We can conclude that the un-binned fitting method is most important 
when fitting a power-law in the tail of a distribution; the binned estimator 
performs nearly as well as the un-binned for the double power-law parameters $\tg$ and 
$\teb$.
The (binned) methods used to report parameters like the ``ankle'' and the ``knee'' 
in \cite{Yamamoto:2007xj} and \cite{Abbasi:2007sv} are sufficient but limited 
by the bin width. 

\newpage

\newpage
\section{Statistical Error: Monte-Carlo Example} \label{sec:MCexe}
\begin{figure}[htbp] 
\Hline
\begin{center}
\includegraphics*[width=110mm, height=60mm]{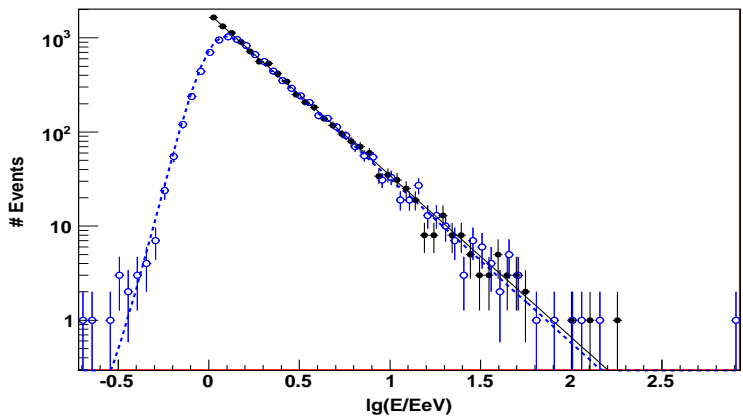} 
\end{center}
\caption{ \label{fig:pplsmear}
  An example of a pure power-law before and after smearing. 
  A histogram of $9000$ events drawn from a single power-law with $\temin = 1.0$EeV and $\tg = 2.75$ is plotted in black filled circles. 
  The best fit (using \equref{equ:lnQ}) power-law for these events is plotted in solid black. 
  The blue open circles are a histogram of these events after being smeared by a 
  Gaussian with variance $0.2E$ (see \secref{sec:MCexe}). 
  The blue dashed curve shows the best fit using \equref{equ:gMmc}.
  To account for the edge effect near $\lg E\sim0$ we use the methods in 
  \secref{sec:StatEE}, namely \equref{equ:gM}.
}
\Hline
\end{figure}

To illustrate the effect the statistical energy smearing has on a pure power-law we 
generate $9000$ MC events from a power-law distribution with $\temin = 1.0$ and $\tg = 2.75$. 
A histogram of these events is represented by the black filled 
circles plotted in \figref{fig:pplsmear}. 
By minimizing \equref{equ:lnQ}, we 
calculate the estimated spectral index for this data to be 
$\tgH = 2.742 \pm 0.019$ (with $\teminH=1.0$, see \secref{sec:GoF}). 
A power-law with these parameters is plotted as the dashed line in \figref{fig:pplsmear}.

To each MC event $E_{i}$ we then add a random number $Y_{i}$ drawn from a 
normal distribution with mean zero and variance $0.2E_{i}$. 
The new events are histogram-ed with blue open circles in \figref{fig:pplsmear}. 
We fit these events by maximizing a likelihood with 
\begin{linenomath*}
\begin{equation} \label{equ:gMmc}
\int^{\infty}_{\temin} \fM(t; \vec{\theta}) \,
                        G(t; \teobs, \sigma_{stat}(t;\vec{p})) \, dt.
\end{equation}
\end{linenomath*}
(compare with \equref{equ:gM}) as the p.d.f. 
and we find that $\tgH = 2.749 \pm 0.020$. The smearing does not effect the estimated 
spectral index, though it does increase the error of the estimate. 
The dashed curve in \figref{fig:pplsmear} shows \equref{equ:gMmc} evaluated at the best fit values. 
Notice that the histogram of the smeared energies deviates from the un-smeared case near $\lg E \sim 0$. 
In \secref{sec:StatEE} we account for this edge effect at the low energy end 
by assuming that the true energies follow the power-law well below the observed minimum energy; 
in constructing the likelihood we choose $0.1\temin$ for the lower rage of integration (compare \equref{equ:gMmc} with \equref{equ:gM}) 
and we re-normalize to ensure a true p.d.f. (see \equref{equ:fMt}). 

\section{Results of CRPropa Toy Set}\label{sec:SumToy}
By applying the statistical tools presented in this paper 
(summarized by steps (\ref{step:BF})-(\ref{step:GoF})
\!\!\!\footnote{Note that since we are not interested in the absolute 
goodness of fit for any of these {\it toy} models to this {\it toy} 
data set, we do not perform step (\ref{step:GoFa}) of \secref{sec:SaC}.}
in \secref{sec:SaC}) to the toy set of $5\times10^{3}$ CRPropa events (see \secref{sec:ats}) 
we illustrate how the tools might be implemented on an actual CR detector.
By construction, this toy set has parameter estimates and, more importantly, 
errors estimates and hypothesis test $p$-values that are numerically comparable with 
those reported by Auger\cite{Yamamoto:2007xj} and HiRes\cite{Abbasi:2007sv}.

In preparation for this paper we generated 14 CRPropa simulations of 
$\sim 2 \times 10^{5}$ events with different injection spectral indexes, 
$\tgin=(2.0,2.1,\ldots,2.6)$, and with different values of maximum generation 
energy, $\temax/\text{EeV}=(400,2000)$.
The (after propagation) 
estimated characteristic break point energy, i.e. $\tehH$ or $\tebH$, is found to be 
independent of the spectral index at the site of generation, $\tgin$.
The estimated spectral index $\tgout$ is found to be linearly related to the input 
spectral index $\tgin$ with linear slope $\sim 1$.
The high energy estimated shape parameters, $\td$ and $\tw$, are more sensitive 
to the maximum generation energy (at the sources) than they are to $\tgin$.

\begin{figure}[htbp] 
\Hline
\begin{center}
\includegraphics*[width=110mm, height=60mm]{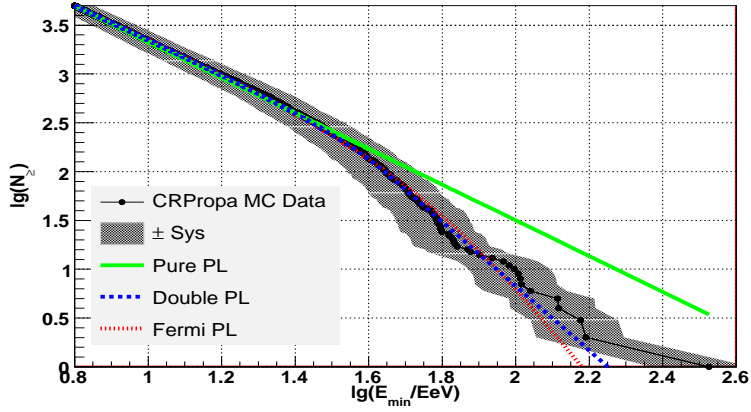} 
\end{center}
\caption{\label{fig:cdf}
  A rank-frequency plot as simulated by $5\times10^{3}$ events from the CRPropa 
  set with parameters $\tgin = 2.6$ and $\temax = 2000$ EeV. 
  For each event (black filled circle) 
  we plot $\lg E$ along the horizontal axis and log-number of events 
  with energy greater than $E$ along the vertical. 
  The models are described in \secref{sec:Models}.
}
\Hline
\end{figure}

\begin{figure}[htbp] 
\Hline
\begin{center}
\includegraphics*[width=110mm, height=60mm]{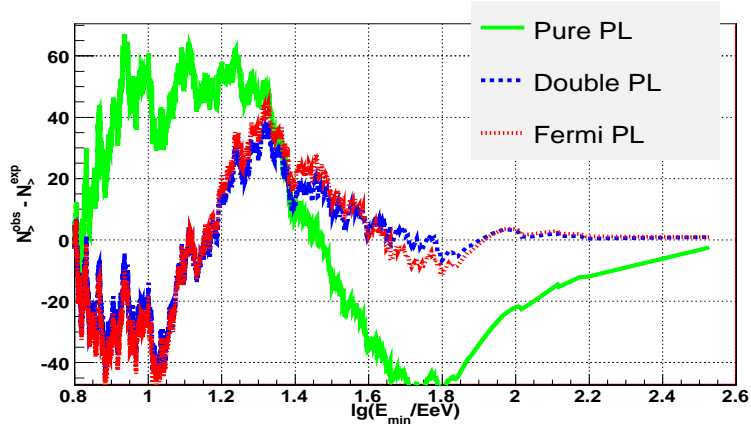} 
\end{center}
\caption{\label{fig:rescdf}
  Using the rank-frequency plot (see \figref{fig:cdf}) we plot the difference 
  between the number of events above a given energy for the toy set  
  $N^{\text{obs}}_{>}$ and that expected by the best fit models $N^{\text{exp}}_{>}$.
  Note that at $\lg \temin/\text{EeV} \sim 1.7$, there are at least forty 
  fewer events observed than 
  expected by the pure power-law fit, i.e. flux suppression.
}
\Hline
\end{figure}

In Figs. \ref{fig:cdf} and \ref{fig:rescdf} we plot the toy data set and the 
best fit models in two (non-binned) ways not commonly seen in the CR literature. 
The first is a {\it rank-frequency} plot. For each event (black filled circle) 
we plot $\lg E$ along the horizontal axis and the log of the number of events 
with energy greater than $E$ along the vertical. 
For each of the models (see \secref{sec:Models}), the vertical axis is 
$\lg (N_{tot}(1 - F(E)))$ where $F(E)$ is the model cumulative distribution function. 
From the rank-frequency plot we derive an instructive visualization tool in \figref{fig:rescdf}; 
we plot the difference between the number of events above a given energy for the toy set  
$N^{\text{obs}}_{>}$ and that expected by the best fit models $N^{\text{exp}}_{>}$.

\begin{figure}[htbp] 
\Hline
\begin{center}
\includegraphics*[width=130mm, height=55mm]{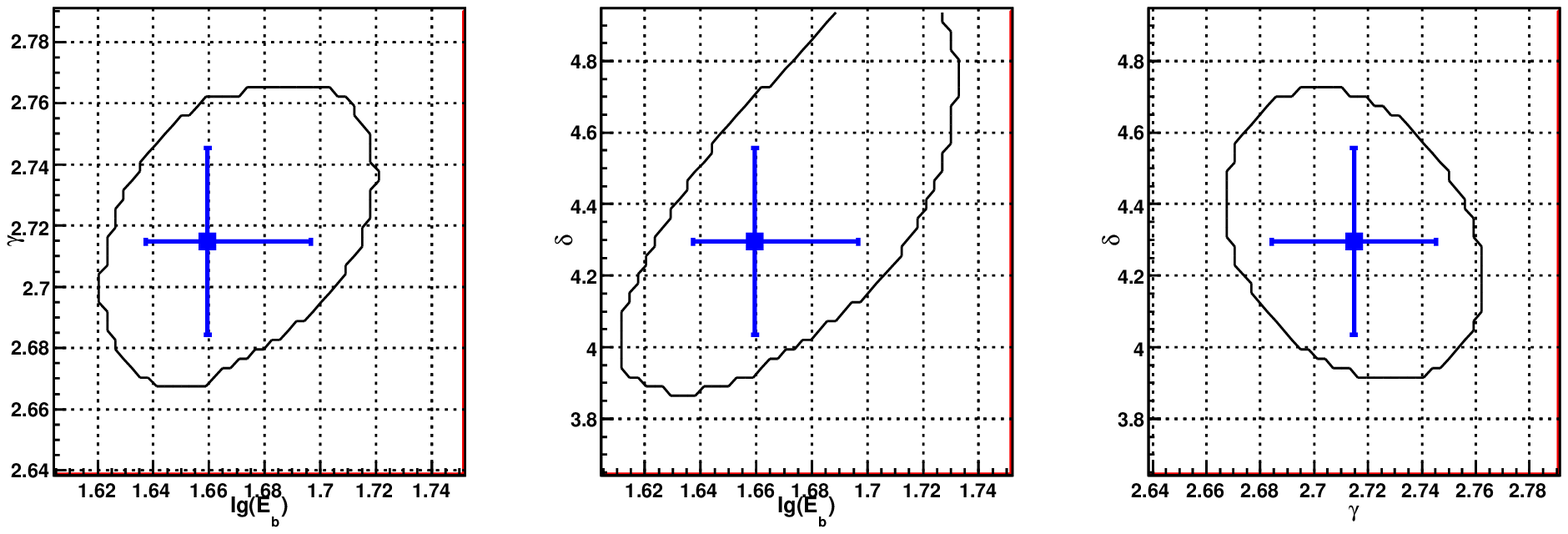} 
\end{center}
\caption{\label{fig:dplLL2}
  The change in log-likelihood $-2\Delta\LDP$ 
  (see \secref{sec:ID}) as a function  
  of the parameters $\tg$, $\teb$ and $\td$ of the double power-law. 
  The data set is the toy set described in \secref{sec:ats}.
  The best estimate for each parameter is plotted as a blue box, 
  the asymmetric one degree of freedom error estimates ($-2\Delta \LDP = 1$) 
  are plotted as solid blue lines and 
  the black contour defines the two degree of freedom error estimate 
  ($-2\Delta\LDP \geq 2.30$).
}
\Hline
\end{figure}

\begin{figure}[htbp] 
\Hline
\begin{center}
\includegraphics*[width=130mm, height=55mm]{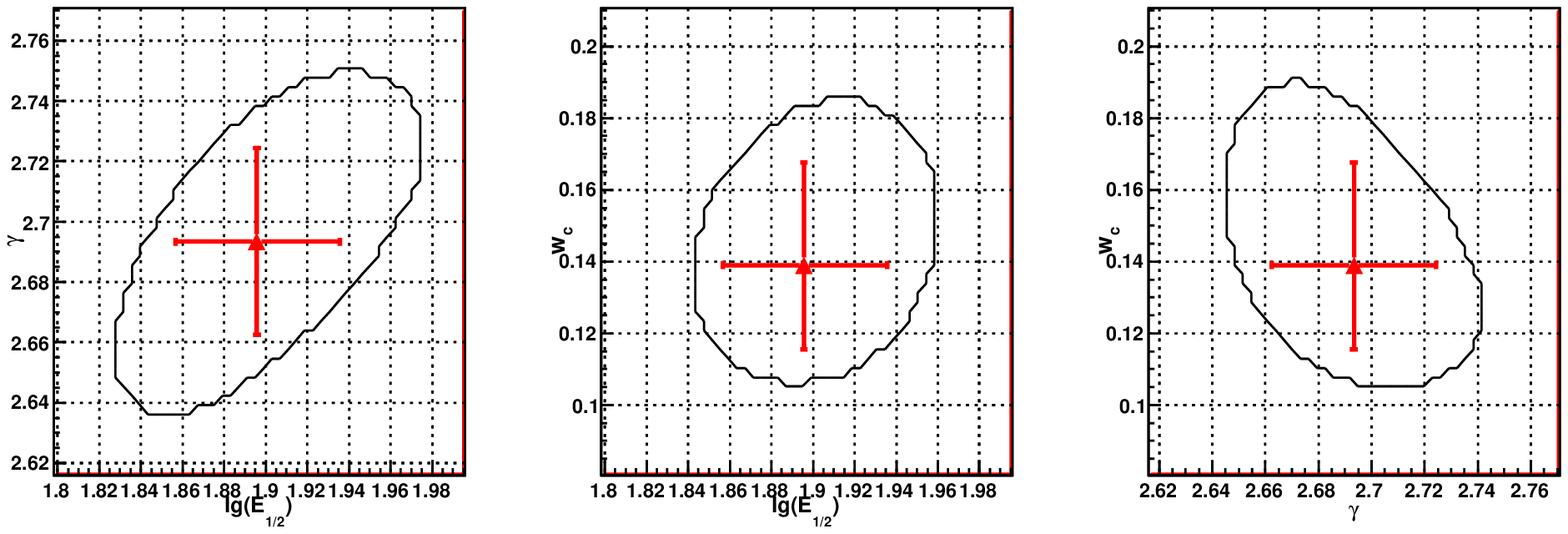} 
\end{center}
\caption{\label{fig:fplLL2}
  The change in log-likelihood $-2\Delta\LFP$ 
  (see \secref{sec:ID}) as a function 
  of the parameters $\tg$, $\teh$ and $\tw$ of the Fermi power-law. 
  The data set is the toy set described in \secref{sec:ats}.
  The best estimate for each parameter is plotted as a blue box, 
  the asymmetric one degree of freedom error estimates ($-2\Delta \LFP = 1$) 
  are plotted as solid blue lines and 
  the black contour defines the two degree of freedom error estimate 
  ($-2\Delta\LFP \geq 2.30$).
}
\Hline
\end{figure}

The best fit pure power-law parameters for the toy set described in \secref{sec:ats} are 
$\temin = 6.31 \pm 0 \pm _{0.82}^{0.82}$ and $\tg = 2.83 \pm _{0.03}^{0.03} \pm _{+0.10}^{-0.07}$ 
where the first error is statistical and the second systematic. 
The tail power significance $\pTP$ is $4.6\sigma$.
The best fit double power-law parameters for the toy set are 
$\temin = 6.31 \pm 0 \pm 0.82$, 
$\tg = 2.71 \pm 0.03 \pm _{+0.10}^{-0.06}$, 
$\teb = 45.7 \pm _{4.1}^{2.3} \pm 9.9$ and 
$\td = 4.30 \pm 0.26 \pm _{+0.20}^{-0.11}$. 
The correlation coefficients are $\rho_{\tg\teb}= 0.18$, 
$\rho_{\tg\td} = -0.15$ and $\rho_{\teb\td} = 0.32$, see \figref{fig:dplLL2}.
The likelihood ratio significance is $\lg p_{\mathcal{R}} = -2.7$.
The best fit Fermi power-law parameters for the toy set are 
$\temin = 6.31 \pm 0 \pm 0.82$,  
$\tg = 2.69 \pm 0.03 \pm _{+0.09}^{-0.06}$,  
$\teh = 78.6 \pm _{7.6}^{6.8} \pm _{19.1}^{18.6}$ and
$\tw = 0.139 \pm _{0.029}^{0.024} \pm _{0.008}^{0.005}$. 
The correlation coefficients are $\rho_{\tg\teh} = 0.61$, 
$\rho_{\tg\tw}$ and $\rho_{\teh\tw} = -0.07$, see \figref{fig:fplLL2}.
The likelihood ratio significance is $\lg p_{\mathcal{R}} = -1.9$.

\newpage

\newpage
\bibliography{libros}
\bibliographystyle{plain}

\end{document}